\documentclass[conference]{IEEEtran}
\IEEEoverridecommandlockouts

\usepackage{pifont}        
\usepackage{cite}
\usepackage{amsmath,amssymb,amsfonts}
\usepackage{algorithmic}
\usepackage{textcomp}
\usepackage{xcolor}
\usepackage{xspace}
\usepackage{booktabs}
\usepackage{graphicx}
\usepackage{tabularx}
\usepackage{multirow}
\usepackage{tcolorbox}
\usepackage{multicol}
\usepackage{enumitem}
\usepackage{fontawesome5}
\usepackage{soul}
\usepackage{titlesec}

\newcommand{\IDEicon}{\faDesktop\xspace}
\newcommand{\CLIicon}{\faTerminal\xspace}
\newcommand{\WEBicon}{\faEdge\xspace}
\newcommand{\SDEicon}{\faSuitcase\xspace}

\newcommand{\TODO}[1][ ]{%
    \ifx#1 %
        \textcolor{red}{TODO}\ %
    \else
        \textcolor{red}{TODO: #1}\ %
    \fi
}
\newcommand{\keyfinding}[1]{
    \begin{tcolorbox}[colback=gray!20, colframe=gray, left=2mm, right=2mm, top=2mm, bottom=1.5mm, boxrule=0mm, sharp corners, leftrule=3pt]
        #1
    \end{tcolorbox}
}
\renewcommand{\keyfinding}[1]{}

\newcommand{\cba}{CBA\xspace}
\newcommand{\cbas}{CBAs\xspace}

\renewcommand{\P}{P}

\def\BibTeX{{\rm B\kern-.05em{\sc i\kern-.025em b}\kern-.08em
    T\kern-.1667em\lower.7ex\hbox{E}\kern-.125emX}}

\begin{document}



\title{
Exploring the Challenges and Opportunities of AI-assisted Codebase Generation 
}

\author{
    \IEEEauthorblockN{Philipp Eibl}
    \IEEEauthorblockA{\textit{Department of Computer Science} \\
    \textit{University of Southern California}\\
    Los Angeles, California \\
    eibl@usc.edu}
\and
    \IEEEauthorblockN{Sadra Sabouri}
    \IEEEauthorblockA{\textit{Department of Computer Science} \\
    \textit{University of Southern California}\\
    Los Angeles, California \\
    sabourih@usc.edu}
\and
    \IEEEauthorblockN{Souti Chattopadhyay}
    \IEEEauthorblockA{\textit{Department of Computer Science} \\
    \textit{University of Southern California}\\
    Los Angeles, California \\
    schattop@usc.edu}
}


\maketitle
 
\begin{abstract}
Recent AI code assistants have significantly improved their ability to process more complex contexts and generate entire codebases based on a textual description, compared to the popular snippet-level generation. These codebase AI assistants (CBAs) can also extend or adapt codebases, allowing users to focus on higher-level design and deployment decisions. While prior work has extensively studied the impact of snippet-level code generation, this new class of codebase generation models is relatively unexplored. Despite initial anecdotal reports of excitement about these agents, they remain less frequently adopted compared to snippet-level code assistants. 
To utilize \cbas better, we need to understand how developers interact with \cbas, and how and why \cbas falls short of developers' needs.
In this paper, we explored these gaps through a counterbalanced user study and interview with (n=16) students and developers working on coding tasks with \cbas. We found that participants varied the information in their prompts, like problem description (48\% prompts), required functionality (98\% prompts), code structure (48\% prompts), and their prompt writing process. Despite various strategies, the overall satisfaction score with generated codebases remained low (mean=2.8, median=3, on a scale of one to five). Participants mentioned functionality as the most common factor for dissatisfaction (77\% instances), alongside poor code quality (42\% instances) and communication issues (25\% instances). We delve deeper into participants' dissatisfaction to identify six underlying challenges that participants faced when using \cbas, and extracted five barriers to incorporating \cbas into their workflows. Finally, we surveyed 21 commercial \cbas to compare their capabilities with participant challenges, and present design opportunities for more efficient and useful \cbas.
\end{abstract}


\section{Introduction}
The widespread adoption of AI-powered coding tools~\cite{so} has caused a paradigm shift in how programmers interact with AI code assistants.
Recent breakthroughs in large language models (LLMs) have allowed these tools to evolve from traditional code-completion to snippet-completion~\cite{copilot}.
This increase in capability has resulted in a popularity explosion, with 62\% of developers reporting using AI-based code tools~\cite{so}.



Recently, growth in the context size of these models~\cite{ding2024longrope}, alongside new reasoning capabilities~\cite{zhu2024llms}, has given rise to another form of code assistant:
codebase-level assistants (\cbas). \cbas create and edit entire codebases using only natural language prompts~\cite{gptengineer}.
Such tools garnered immediate interest in the developer communities, as they promised functionality at the repository and project level. Some of these models went further to present themselves as ``artificial software engineers'' \cite{devin}. Despite promising capabilities, \cbas have not seen the same adoption as their snippet-level peers, such as GitHub Copilot.
While the snippet-level code assistants have garnered significant research interest in various areas of SE literature, from productivity~\cite{prod} and usability~\cite{liang2023usability} to error analysis~\cite{liu2023your}, \cbas have remained vastly understudied.
To understand how to build \cbas that can live up to their promised potential, we need to study where and how \cbas currently fall short for developers. In this paper, we investigate the usability, challenges, and opportunities of this new generation of coding agents. To this end, we address the following research questions:

\textbf{RQ1. How do programmers prompt \cbas?}
Prompts serve as the main interface between the user and the tool. To understand how programmers interact with the \cbas, we analyze the types of information users (developers and programmers) include (RQ1a) and how their prompting behavior evolves during composition (RQ1b).

\textbf{RQ2. How well do \cbas satisfy users?}
With the various prompting strategies, we examine the distribution of users' satisfaction with the generated codebases. We delve deeper into what factors influence satisfaction among users.

\textbf{RQ3. What are the challenges of using \cbas for codebase generation?}
With the low satisfaction rate of \cba codebases, we investigate the underlying technical and interaction-related challenges that users face when using \cbas. 


\textbf{RQ4. What are the barriers to integrating \cbas into developer workflows?} Finally, we investigate what users perceive as broader adoption barriers that raise usability concerns and distrust among users, further hindering \cba integration into their development workflows.

To answer these questions, we conducted a counterbalanced user study with 16 users (eight software developers and eight students), followed by a reflective interview and survey.
Participants were asked to use either GitHub Copilot or GPT-Engineer to complete three code tasks (e.g., create or edit a repository) while thinking aloud.
We qualitatively and quantitatively analyzed their prompting process, interaction patterns, satisfaction factors, and perceived challenges.

We observe substantial variation in how developers write prompts, varying the information within a prompt and the focus---from visual layout to problem statements to functionality---often, omitting details like requirements and tests (RQ1). Our participants were frequently dissatisfied with the generated code, with only about 50\% of the outputs meeting their expectations. Participants perceived that this dissatisfaction stemmed from \cba generated code with missing functionality, execution failures, and lack of clarity or comments in the codebase (RQ2). Based on participant accounts, we then identified the challenges they faced when working with \cbas, like dealing with missing and blank code, bidirectionally inadequate communication protocols, and rectifying ignored requirements (RQ3) and broader barriers to adoption of \cbas (RQ4).

Finally, to examine the mismatch between challenges and barriers participants faced and the capabilities offered by current \cbas, we surveyed the features provided by 21 \cbas. We identify seven core capability dimensions offered by these tools. While a handful of these capabilities mitigate some of the user challenges and barriers, we discuss further design opportunities for future \cbas to address user needs and generate satisfactory codebases. These recommendations point toward more supportive prompting, increased transparency, and a collaborative codebase generation process. 



\section{Related Work}

We discuss prior work on AI-assisted programming, focusing on usability and developer experiences.

\textbf{Usability of AI Coding Assistants.}
A growing body of work investigates how developers use AI coding tools and the challenges they face. Liang et al. surveyed developers and found that while assistants reduce  effort and boost productivity, they often fail to meet functional or non-functional requirements~\cite{liang2023usability}. Other studies echo similar sentiments: developers find assistants helpful for exploration and acceleration, but struggle with correctness, context awareness, and maintaining control~\cite{barke_grounded_2022,devNeeds}. Pinto et. al. found that contextualized assistants can aid internal documentation use but still suffer from a lack of variable responses and code reuse~\cite{pinto2023developerexperiences}. Others observed usability challenges in accessibility~\cite{codea11y}, developer intent understanding~\cite{sourceCodeCompletion}, and input formulation~\cite{syntax}.

\textbf{Human--AI Interaction in Coding.}
Prior work examined how programmers engage with AI-generated code, identifying desired delegation areas (e.g., tests, comments) and recommending frameworks for understanding interaction patterns~\cite{betweenLines,sergeyuk2025,sadra-trust}. Studies also highlight how LLM-based tools impact trust, autonomy, and code review dynamics~\cite{takingFlight}.

\textbf{Codebase-Level Generation Agents.}
While snippet-level assistants (e.g., Copilot) have seen widespread adoption and study, repository-level assistants remain less explored. Jiang et al.~\cite{jiang2024survey} outline core challenges such as handling cross-file dependencies and exceeding context limits. Commercial and open-source tools like GPT-Engineer, Devin, Cursor, and Aider claim support for project-scale tasks, yet little work systematically evaluates their usability or real-world effectiveness in developer workflows. Some studies benchmark model performance~\cite{jimenez2023swe}, but empirical research on how developers interact with \cbas is scarce.

In contrast to technical efforts focused on optimizing model inputs~\cite{stallplus,zhang2024hcp,repofusion}, our work explores how developers prompt, evaluate, and adopt \cbas surfacing gaps in alignment, communication, and workflow fit.

\section{Methodology}
\begin{figure}
    \centering
    \includegraphics[width=\linewidth]{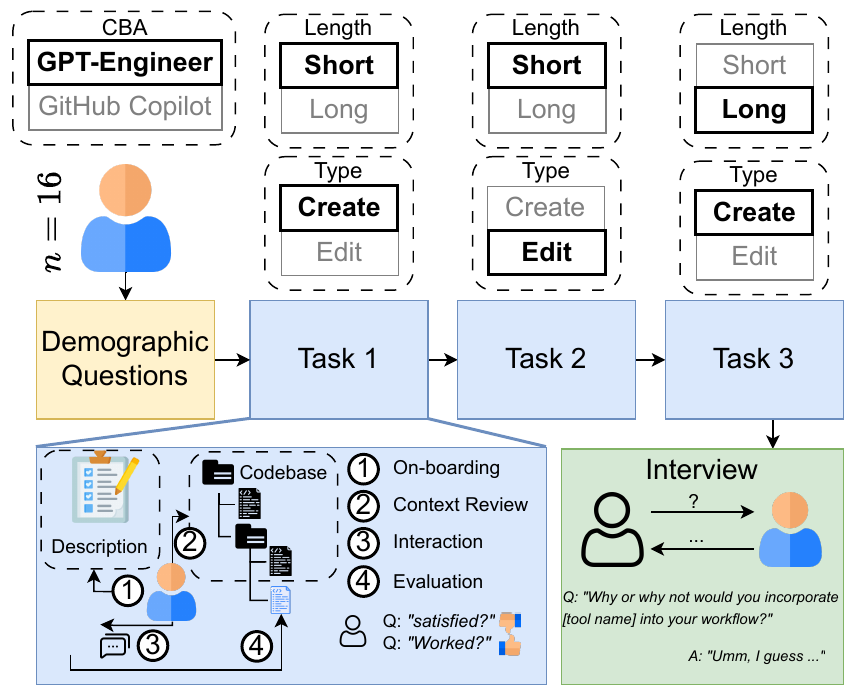}
    \caption{An overview of our methodology. Each participant was assigned to a \cba and completed three coding tasks. For each task, participants first received a task description (1). If the task was to edit a codebase, participants first reviewed that codebase (2). Participants then prompted the \cba to generate or edit the codebase (3) and evaluated the resulting code (4). After each task, they answered a set of questions about the interaction. After the final task, participants were interviewed about their experience.}
    \label{fig:method-overview}
\vspace{-1em}
\end{figure}
\noindent We ran a user study (n=16) in which participants completed create or edit tasks on small to medium codebases. We logged their prompts, interaction patterns, and satisfaction ratings to uncover behaviors and drivers of their experiences (Fig. \ref{fig:method-overview}).

\subsection{Recruitment}
We recruited 16 adult participants (6F/10M) who were US residents and fluent English speakers using convenience and snowball sampling (referred to as P-i henceforth).
\P1-\P8 were graduate student programmers, and \P9-\P16 were professional developers from seven companies. 
This study was approved by our institution's Institutional Review Board.
\label{sec:method.participants}
\renewcommand{\arraystretch}{1.2}  

\begin{table}[ht]
\centering

\caption{
\textbf{Participant Demographics}. Sections \ref{sec:method.participants} and \ref{sec:method.study-design} contain explanations for the abbreviations used in the column names.
}

\resizebox{\linewidth}{!}{
\begin{tabular}{c c c c |cccc}
\hline
\multirow{2}{*}{\textbf{PID}} & \multirow{2}{*}{\textbf{XP}} & \multirow{2}{*}{\textbf{AI use}} & \multirow{2}{*}{\textbf{\cba}} & \multicolumn{4}{c}{\textbf{Task}} \\ \cline{5-8}
& & & & \textbf{SC} & \textbf{SE} & \textbf{LC} & \textbf{LE} \\ \hline
1  & 0.5 & Day to day     & Copilot       & \checkmark & \checkmark & \checkmark &           \\ \hline
2  & 2   & Once or twice  & Copilot       & \checkmark & \checkmark &           & \checkmark \\ \hline
3  & 4   & Occasionally   & Copilot       & \checkmark &           & \checkmark & \checkmark \\ \hline
4  & 1   & Never          & Copilot       &           & \checkmark & \checkmark & \checkmark \\ \hline
5  & 1   & Day to day     & GPT-Engineer  & \checkmark & \checkmark & \checkmark &           \\ \hline
6  & 0.5 & Occasionally   & GPT-Engineer  & \checkmark & \checkmark &           & \checkmark \\ \hline
7  & 3   & Occasionally   & GPT-Engineer  & \checkmark &           & \checkmark & \checkmark \\ \hline
8  & 1   & Never          & GPT-Engineer  &           & \checkmark & \checkmark & \checkmark \\ \hline
9  & 0.5 & Never          & Copilot       & \checkmark & \checkmark & \checkmark &           \\ \hline
10 & 6   & Day to day     & Copilot       & \checkmark & \checkmark &           & \checkmark \\ \hline
11 & 1   & Day to day     & Copilot       & \checkmark &           & \checkmark & \checkmark \\ \hline
12 & 2   & Day to day     & Copilot       &           & \checkmark & \checkmark & \checkmark \\ \hline
13 & 2   & Occasionally   & GPT-Engineer  & \checkmark & \checkmark & \checkmark &           \\ \hline
14 & 2   & Once or twice  & GPT-Engineer  & \checkmark & \checkmark &           & \checkmark \\ \hline
15 & 5   & Once or twice  & GPT-Engineer  & \checkmark &           & \checkmark & \checkmark \\ \hline
16 & 3   & Day to day     & GPT-Engineer  &           & \checkmark & \checkmark & \checkmark \\ \hline
\end{tabular}
}
\label{table:demographics}
\end{table}
Table~\ref{table:demographics} summarizes the study-related characteristics of our participants.
\emph{AI use} indicates how often participants used AI coding assistants in the past year,
and \emph{XP} denotes years in their current role, ranging from six months to six years.
Participants’ JavaScript proficiency ranged from 1 to 5 (median = 2.5); Python from 2 to 5 (median = 4).

\subsection{Study Design}
\label{sec:method.study-design}
\subsubsection{Coding Session}
We observed participants while they completed assigned tasks using a \cba, followed by a brief interview.
The median session time was 65 minutes, with the shortest and longest sessions taking 48 and 85 minutes.
Participants' assignment was counterbalanced between their \emph{Role}, \cba treatment they will use for the tasks (\emph{CBA}), and the \emph{Task Type}. Table~\ref{table:demographics} shows the \cba and task assignment across participants.
\emph{CBA} indicates the codebase assistant used during the study, and \emph{Task} identifies the types of tasks performed.

\textbf{Treatment (\cbas).}
Participants were assigned to two \cba treatments: Copilot and GPT-Engineer (using gpt-3.5-turbo as the underlying LLM).
Half of the participants completed all the tasks using GPT-Engineer, while the other half completed them using GitHub Copilot.
Each participant used only one tool throughout the session to maintain within-subject consistency.
We did not allow the use of GPT-Engineer's clarification feature, as no equivalent exists for Copilot.

\textbf{Tasks.}
Each participant completed three code tasks within their assigned treatment group. 
We varied the tasks across four categories based on two dimensions: 
task length and task type. 
This yielded four distinct task groups:
\begin{itemize}[leftmargin=*, noitemsep]
    \item Short-Create (SC): Create a small codebase (1–2 files)
    \item Short-Edit (SE): Edit a small existing codebase
    \item Long-Create (LC): Create a larger codebase (3+ files)
    \item Long-Edit (LE): Edit a larger existing codebase
\end{itemize}
Each participant performed at least one create, one edit, one short and one long task.
Task assignment was also counterbalanced across role and \cba (see Table~\ref{table:demographics}).
For each task category (SC, SE, LC, LE), we designed two tasks to mitigate learning effects.
For instance, in the SC group, one task required creating a simple calculator application, and the other a simple stopwatch. 
Task descriptions and code were crafted by the authors and are in the supplementary material~\cite{supp}.

All the tasks followed the below sequential process:
\begin{enumerate}[leftmargin=*]
    \item On-boarding: Participants were shown a task description and on-boarded with the task objectives.
    \item Context Review (Edit tasks only): Participants were given time to explore the existing codebase before prompting.
    \item Interaction: Participants were asked to submit a prompt addressing the given task objectives.  
    \item Evaluation: Participants attempted to run the code and rated the generated code’s functionality, quality, and completeness providing justifications during or after the interaction.
\end{enumerate}

All participants performed the study in the same development environment (Visual Studio Code) hosted on one of the authors' machines to mitigate any confounding effects from the development environment.
We recorded audio and screen video during the sessions using Zoom.

The exact prompts for the tasks can be found in ~\cite{supp}.
After each task, we asked participants to rate the generated code from 1-5 and to explain their rating.
\subsubsection{Interview}
After the coding tasks, we conducted a semi-structured interview, with questions about the participant's interaction and experience with the tool.
The full interview protocol is available in the supplementary material ~\cite{supp}. Interviews were audio recorded and transcribed for qualitative analysis.
\subsection{Analysis}
We analyzed participant behavior, interaction patterns, and codebase evaluations to address our research questions.

\subsubsection{Prompt Content (RQ1a)}
\label{sec:method-analysis-rq1a}
To examine prompt contents, two authors conducted open inductive coding with negotiated agreement on all 48 prompts (16 participants × 3 tasks). The resulting codebook included functional requirements, GUI descriptions, control flow, technical specifications, and other metadata. One author applied the codes, resolving ambiguities collaboratively. We then manually grouped prompts thematically by information content.

\subsubsection{Prompt Writing Process (RQ1b)}
We annotated each session using three interaction types: writing, editing, and pausing ($\ge 0.5$s). One author used screen recordings to time-code all 48 tasks and manually clustered them by visual inspection to identify prompt authoring styles.

\subsubsection{Rating Justifications (RQ2)}
\label{sec:method-analysis-rq2}
To identify reasons behind codebase ratings, we inductively coded 20\% of transcripts, refining codes over 11 rounds ($\kappa=0.87$, $p<0.001$). We grouped final codes into 10 themes (e.g., functionality, usability, executability, correctness, completeness, guidance) and applied them to the full dataset.

\subsubsection{Challenges of using \cbas (RQ3)}
We examined mismatches between participant expectations and generated codebases by analyzing negative evaluations, generated code, and error outcomes. Two researchers performed thematic analysis, identifying six common challenges: missing code, unusable output, inadequate communication, ignored requirements, ignored context, and missing instructions.

\subsubsection{Workflow Integration Deterrents (RQ4)}
To uncover adoption barriers, we inductively coded the final interview responses, from which five socio-technical deterrent themes emerged.

\section{\textbf{RQ1}. How do programmers write \cba prompts?}
We examined both the content of prompts (RQ1a) and the writing process (RQ1b) to understand how developers compose prompts for \cbas.

\subsubsection*{\textbf{[RQ1a]} What information do users include in their prompts?}
\begin{table}[t]
\centering
\caption{Distribution of Included Information}
\begin{tabular}{lrrr}
\toprule
Information Type & Mean & Count & Corr. \\
\midrule
Functional Requirements & 0.98 & 47 & 0.09 \\
Problem Description & 0.56 & 27 & 0.09 \\
GUI Description & 0.48 & 23 & -0.19 \\
Flow/Interactivity & 0.48 & 23 & -0.06 \\
Constraints & 0.31 & 15 & -0.19 \\
Checklist & 0.23 & 11 & -0.22 \\
Implementation Guidance & 0.23 & 11 & -0.37* \\
Error Handling Spec. & 0.19 & 9 & -0.04 \\
Reusability/Extensibility & 0.08 & 4 & -0.02 \\
Familiar Examples & 0.06 & 3 & -0.07 \\
Testable Examples & 0.06 & 3 & -0.20 \\
File/Modular Structure & 0.12 & 6 & 0.15 \\
\midrule
Tone: Imperative/Command & 0.83 & 40 & 0.31* \\
Tone: Suggestive & 0.08 & 4 & -0.14 \\
Tone: Mixed & 0.08 & 4 & -0.28* \\
\midrule
Level of Detail: High & 0.38 & 18 & 0.09 \\
Level of Detail: Medium & 0.46 & 22 & -0.02 \\
Level of Detail: Low & 0.17 & 8 & -0.09 \\
\bottomrule
\end{tabular}
\label{tab:prompt_info}
\end{table}
Participants included a wide variety of content types in their prompts, reflecting varying mental models of how \cbas work.
Prompts were often a blend of natural language descriptions, implementation constraints, interface-level detail, and behavioral flow.
Users didn’t just describe what the program should do, but also how it should operate or behave during use.

Table~\ref{tab:prompt_info} shows the different types of information participants included in their prompts. 
Despite the diversity in types, users consistently structured their prompts around functional requirements (n=47) and problem description (n=27), with many adding interactivity constraints (n=23) or GUI descriptions (n=23). Some participants added functional and behavioral constraints (n=15), a checklist of intended behaviors (n=11), and additional implementation guidance like packages and libraries to use (n=11). However, error handling (n=9) and test cases (n=3) were rarely included. This highlighted a consistent under-specification of criteria for robustness.

We also observed that participants mostly structured their prompts as direct commands, using an imperative tone (n=40) instead of making suggestions. They also included medium (n=22) or high (n=18) amounts of detail in their prompts to ensure the codebase generated will satisfy their needs. 

The \emph{‘Corr.’} column shows the Pearson correlation (significant correlations marked with an asterisk) between the presence of a certain prompt feature and satisfaction score with the generated codebase (from 1 to 5 inclusive).
Prompts written in a highly imperative style (e.g., “Make sure the button does...”) had a positive correlation with satisfaction (r = 0.3). While not high, this could suggest that \cbas respond better to clear, directive language due to the similarity of the imperative tone of examples of the instruction-tuning stage~\cite{kang2024unfamiliar} of the LLM. Whereas, prompts that included implementation guidance, e.g., “use a PUT route” or “call fetchOrders”, had the highest negative correlation with satisfaction (r = -0.3).

Inspecting co-occurring information types, we identified three prevalent types of prompts based on information content:

\textbf{Type I: Interface/Interaction focused prompts}
[P1, P3, P5-P16]
14 prompts focused heavily on GUI layout, user interaction flow, or dynamic behavior. 
For instance, one user wrote, “In the middle of the page will be a large box, aka the chat log \ldots there will also be a check box on the left-hand side of the text input box that indicates which user is talking” [P5]. 
These prompts were rich in user-facing functionality and often described the sequence of interactions or component layout in detail, sometimes even without naming specific frameworks or programming paradigms. 
    
\textbf{Type II: Behavior-focused prompts}
[P2, P4-P7, P9-P11, P14]
8 prompts were grounded in the description of a problem domain without visual or structural specifics. 
For instance, P7 prompted, “Create a to-do list for the tasks I need to do in March... the to-do list should mark the status if I have finished the task”. 
These prompts capture user goals clearly but leave many implementation details open, including technical architecture or user interface. 
    
\textbf{Type II: Functionality-focused prompts}
[P1, P4, P5, P12-P14]
6 prompts were tightly scoped and often limited to direct feature requests, sometimes resembling patch-level updates or code-level edits. 
For instance, ``Add the operation ‘exponential’, which takes two numbers and calculates the first number to the power of the second number'' [P1].
These prompts often assume a shared context with the system, implicitly referencing existing code structures or expected functionality.

A majority of users did not describe the underlying code or architecture they imagined.
Instead, they focused on surface-level interactions or user flows. 
This absence of “code-oriented prompts” suggests that developers, even when coding, think in terms of interface logic instead of implementation.


Overall, users do not appear to begin by mentally constructing the codebase they want.
Rather, they articulate their needs in terms of problem scenarios, behavioral outcomes, or interaction flows.
This orientation may be well-suited to high-level generation but also causes mismatch or incompleteness.


\subsubsection*{\textbf{[RQ1b]} What is the prompt writing process?}
We found that participants spent 40\% of the time writing the prompt, 15\% editing the prompt, and did not interact with the keyboard 45\% of the time. 
However, prompts were rarely written in a single pass; participants oscillated between these three actions, with a mean count of 24.5 actions per prompt.
Participants generally maintained a consistent prompting process over time.
The variability in their action count and action duration was lower within individuals ($\sigma_\text{count}^\text{w} = 7.1$, $\sigma_\text{duration}^\text{w} = 5.3$) than between different participants ($\sigma_\text{count}^\text{b} = 10.2$, $\sigma_\text{duration}^\text{b} = 8.6$). 
Based on Welch's t-test, both differences were statistically significant with $p=0.01$ ($t=2.73$) for duration and $p=0.04$ ($t=2.21$) for count differences.

However, we found that participants' actions varied with task length.
While they spent about one additional minute on longer tasks 
($\mu_{\text{prompt}|\text{long}} = 2\,\text{m}30\,\text{s}$)
compared to short tasks
($\mu_{\text{prompt}|\text{short}} = 1\,\text{m}36\,\text{s}$), further inspection revealed that participants spent a shorter fraction of the time pausing during longer tasks ($\mu_{\text{\%pause}|\text{long}} = 0.38$ vs. $\mu_{\text{\%pause}|\text{short}} = 0.51$), whereas spent more time writing and editing during long tasks. Although these differences were not statistically significant, they captured a general picture of the prompting process. 


Examining the sequences of these actions to identify meaningful patterns, we found that participants' prompting patterns reflect their problem-solving process. For instance, in 10 instances, participants [P5, P4, P13, P15, P16] paused for longer periods before writing continuously for a while. This pattern indicated a \textbf{Planning} phase, where the pause was followed by writing detailed information about functionalities or behavior. We also observed four instances [P2, P9, P12] where participants often went back and forth between short blocks of writing and editing (less than 5 seconds on average), potentially with thinking breaks in between. This pattern aligned with \textbf{Exploring/Tinkering} phase, where participants explore ideas while writing the prompt, but change their minds later. In another 5 instances [P1, P13], participants followed a steady writing action with few or no edits, writing continuously for more than 10 seconds on average, and spent less than 15\% of the time without writing. This process suggested that participants had a clear idea of how to construct their prompts and were \textbf{Executing} their idea through writing.

\keyfinding{
Prompting process was iterative: users frequently switched between writing, editing, and pausing. Users were consistent in their prompting process. The sequences of the actions in the prompting process reflected the users' problem solving process, prompting patterns indicated when users were planning, exploring, and executing their plan.
}

\section{\textbf{RQ2}. How well do \cbas satisfy users?}
\label{sec:rq2}

\begin{figure}
    \centering
    \includegraphics[width=1\linewidth]{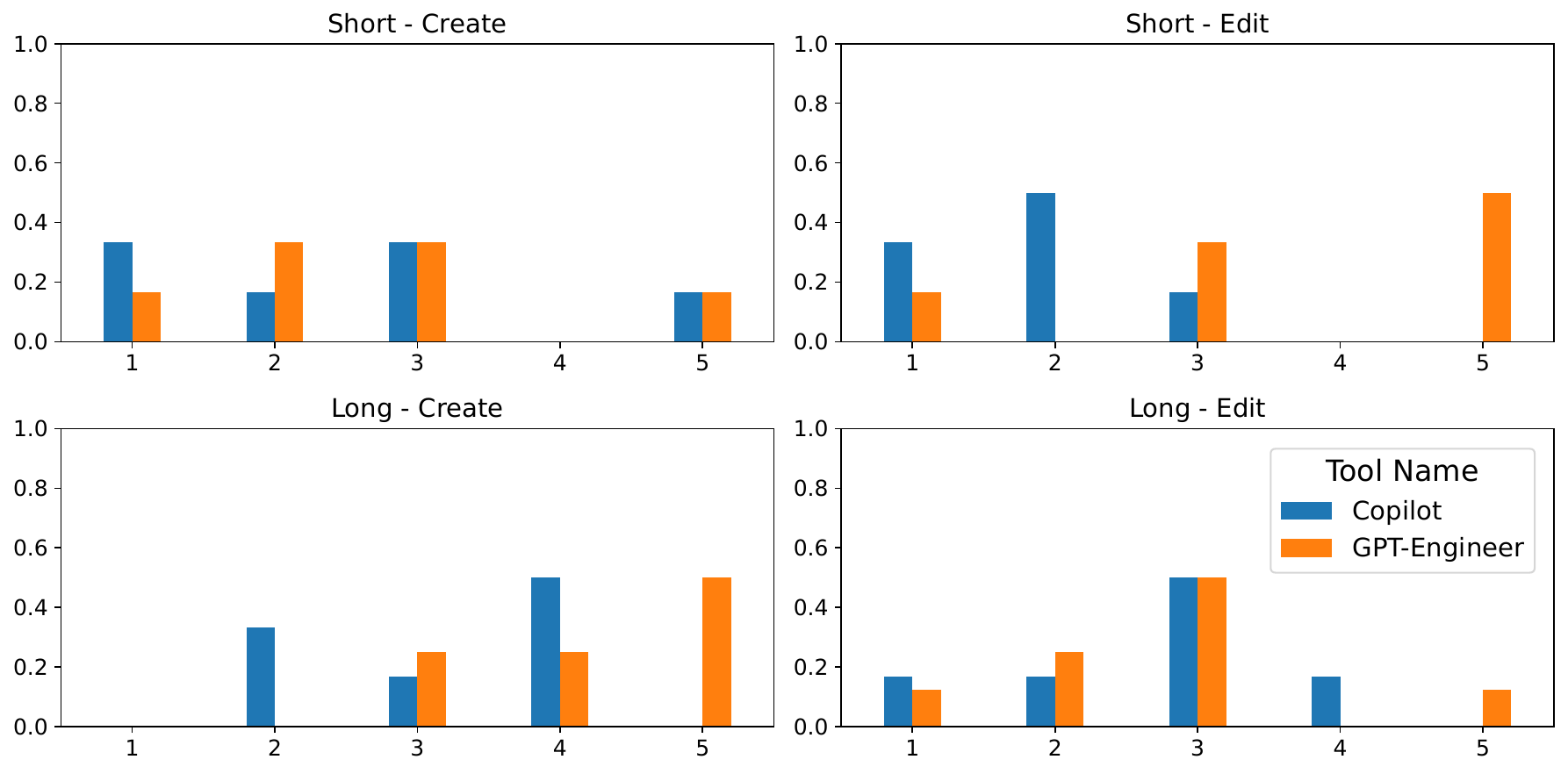}
    \caption{Distribution of satisfaction scores by task type, task length, and \cba. The X-axis represents satisfaction scores; the Y-axis shows the fraction of participants assigning each score.}
    \label{fig:satisfaction-rates}
\vspace{-1.5em}
\end{figure}

Prompt structure alone does not determine a successful outcome.
We observed that even well-structured prompts led to unsatisfactory code.
To better understand the challenges users face with current \cbas, RQ2 investigates when and how the outputs meet user expectations or fail to do so. Overall, participants' mean satisfaction ratings across generated codebases were 2.8 (s.d. = 1.28, median = 3) on a scale of 1-5.

We further examined the satisfaction ratings with respect to the different tasks and \cbas. Figure~\ref{fig:satisfaction-rates} shows the distribution of satisfaction ratings across the task types and task lengths, separated by the two treatment \cbas. These plots share the finding that overall satisfaction gravitates towards the middle of the scale, with the exception of \emph{short-edit} tasks with a slightly higher overall satisfaction. A chi-square test of independence (appropriate for such a sample sizes~\cite{roscoe1971investigation}) revealed that the proportion of satisfaction distribution across the task settings and \cbas is independent ($\chi^2 = 2.74, p=0.43$). Overall, participants were demonstrated unremarkable satisfaction with the generated codebases.

To understand what factors informed the participants' satisfaction, we conducted a thematic analysis by unitizing participant comments and applying open coding to recurring themes.
Table \ref{tab:factors} lists total counts for each factor participants mentioned during their evaluation of the code. 
\emph{\#Mentions}, \emph{\#Tasks}, and \emph{\#Participants} refer to the total number of times that a factor was mentioned, the number of tasks in which the factor was mentioned, and how many participants mentioned it. The table also shows the satisfaction ratings associated with each factor and the Pearson correlation coefficient.

\renewcommand{\arraystretch}{1.15}  
\begin{table}[t]
\centering
\caption{Factors. Counts format is \#Mentions/Tasks/Participants.}
\vspace{-1em}
\resizebox{\linewidth}{!}{
\begin{tabular}{l l r l r}
\toprule
\textbf{Factor} & \textbf{Counts} & \textbf{Sub-Factor} & \textbf{Counts} & \textbf{Corr.} \\
\midrule

\multirow{2}{*}{Functionality} & \multirow{2}{*}{25/21/12} & Unmet func. req. & 16/13/9 & -0.02 \\
& & Met func. req. & 9/9/7 & 0.34* \\

\hline
\multirow{2}{*}{Usability} & \multirow{2}{*}{10/9/7} & Usable code & 7/7/6 & -0.07 \\
& & Unusable code & 3/3/3 & -0.20 \\

\hline
Executability & 9/9/7 & Non-executable code & 9/9/7 & -0.24* \\

\hline
\multirow{2}{*}{Prompt Specificity} & \multirow{2}{*}{7/7/6} & Detailed prompt & 2/2/2 & -0.14 \\
& & Vague prompt & 5/5/4 & 0.19 \\

\hline
\multirow{2}{*}{Correctness} & \multirow{2}{*}{7/7/6} & Incorrect code & 5/5/5 & 0.02 \\
& & Correct code & 2/2/2 & 0.06 \\

\hline
Completeness & 7/7/6 & Incomplete code & 7/7/6 & -0.10 \\

\hline
\multirow{2}{*}{Readability} & \multirow{2}{*}{7/6/5} & Unreadable code & 5/5/4 & -0.00 \\
& & Readable code & 2/2/2 & 0.00 \\

\hline
Maintainability & 6/6/6 & Poor maintainability & 6/6/6 & 0.08 \\

\hline
\multirow{2}{*}{Guidance} & \multirow{2}{*}{5/5/4} & Missing guidance & 4/4/4 & 0.03 \\
& & Good guidance & 1/1/1 & 0.02 \\


\bottomrule
\end{tabular}
}
\label{tab:factors}
\vspace{-2em}
\end{table}

\noindent\textbf{Functionality.}
Participants frequently remarked on unmet functional requirements. In 16 of the 48 tasks (33\%), they noted functionality issues—either missing features (e.g. \P6: “it's not showing me the inventory”) or features that diverged from expectations (e.g., \P16: “I was thinking slightly different as in human readable”). These shortcomings ranged from partial gaps (\P1: “doesn't actually print the time”) to entirely ineffective outputs (\P9: “it didn't accomplish anything”).
Despite this, unmet requirements showed only a weak correlation with satisfaction ($r = -0.02$).
By contrast, mentions of met functional requirements--whether partial (e.g., \P15: “edit-inventory symbol was correct”) or complete (e.g., \P14: “met all of the requirements”)--had the strongest positive correlation with satisfaction ($r = 0.45$, $p = 0.0015$).

\noindent\textbf{Usability.}
In nine tasks (19\%), participants raised the usability of the generated code as a factor in their decision to use or not use \cbas. Some participants shared positive feedback on the generated code's usability. \P4 remarked that the generated code was ``a starting point" which they ``can build on (\P4)." With additional positive reflections, \P15 said, ``[The CBA] gave you the structure and the boilerplate for some of the things".
However, negative feedback was also presented in three tasks.
It occurred over a spectrum of frustration, from participants saying they ``will have to make a lot of changes" (\P1) to others describing the codebase as entirely unusable. 
The latter was exemplified by \P9's commenting, ``I would trash the entire thing and try to start over. I would never use this.''

\noindent\textbf{Executability.}
Nine participants expressed that code that is not ``running off the bat'' (\P9) was negatively impacting by their satisfaction.
We found that participants consider ``executability" an important dimension, highlighted by the factor's Pearson correlation of $-0.24$ ($p=0.004$) with satisfaction.
Notably, some types of programs were ``executable" at higher rates than others, as, for example, web-based code bases were typically ``executable" despite erroneous code.
\\
\noindent\textbf{Prompt specificity.}
In seven tasks, participants also adjusted their expectations to the perceived detail they added to the prompt. 
In the majority (5/7) of instances, the participants were lenient with missing functionality because of a perceived lower degree of detail in the prompt. In discussing the poor functionality of their generated codebase, (\P2) reflected, ``Due to lack of specificity that I provided in the initial prompt, this makes sense".
However, the participants' expectations rose as they felt their requirements had been ``expressed very clearly [yet the \cba did not generate satisfactory code]" (\P10).
\\
\noindent\textbf{Correctness.}  
Participants commented on correctness in both positive and negative terms, focusing on whether the generated code was logically or syntactically correct. In two tasks, participants praised specific aspects as correct—e.g., \P1: ``The logic itself is fine,'' and \P11: ``The arguments [...] are exactly correct.'' In three cases, they noted alignment with their own approach—e.g., \P6: ``It did everything that I would have done.''

Negative comments, more frequently, pointed to syntax or logic errors, such as incorrect formatting (\P8: ``has an equals and then has the semicolon, which is not correct'') or flawed logic (\P10: ``This is not the correct start time'').
Finally, we also observed statements about generally ``buggy" code (\P15).
\\
\noindent\textbf{Completeness.}
Incomplete code was a notable detractor of satisfaction which raised in six sessions. 
This included cases where parts of the code were missing, undefined, or uninitialized including cases of missing variables or attributes (\P5, ``\texttt{.log\_history} isn't even defined"), to missing functions (\P9, ``there's no evaluate function"), to missing files (\P8, ``because we don't have that package.json [file], which is really important.").
We further observed two cases where code was left ``as an assignment to the reader" (\P15) such that the \cba ``didn't create everything [completely]" and instead ``just did some comments" (\P8) or docstrings.
In some instances, usage of code that was references to undefined functions which confused the participant, as was the case for \P9 when they found ``no \texttt{handle\_math\_error()} function. So I don't know where that came from."
\\
\noindent\textbf{Readability.}
Of all seven readability-related statements, five talked about the scarcity of comments, mentioning they ``would like more comments on the code" (\P12), while four talked about readability in general.
The other two stated that the documentation increased their satisfaction with the code.
\\
\noindent\textbf{Maintainability.}
Participants made negative remarks on the maintainability of the code in six instances, such as when \P9 pointed out that ``maintenance [of this code] in the long term could be a problem".
Participants unhappy with the structural aspects of the code expressed that it was ``duplicates and [was] breaking encapsulation (\P15)".
Within the editing tasks, we observed statements related to the new code's divergence from the existing code. 
For example, they did not like it when the \cba did not ``use the function that is already there, and just call it (\P12)".
A representative statement for this type of factor was made by \P10: ``the fact that it decided to give me a new function in favor of my old functions is weird."
Finally, participants disliked it when they saw methods/functions ``that didn't seem to be used in the code. (\P4)" As mentioned by \P15: ``It has an unused import. So that's annoying."
\\
\noindent\textbf{Guidance.}
Missing guidance's negative effect on participants' satisfaction manifested exclusively in editing tasks and varied by \cba because while GPT-Engineer would edit source code directly, Copilot would not apply code changes itself but instead produce code and provide the user with a description of where to place that code in natural language.
For GPT-Engineer, participants sometimes did not know about ``what changes have been made as compared to the previous code (\P6)." 
While for Copilot, participants sometimes did not ``know where to put the first chunk of code (\P3)."
Thus, both strategies can cause frustration, and it is not clear that either option results in higher satisfaction.

\section{\textbf{RQ3}. What are the challenges of using \cbas for codebase generation?}
By observing patterns in dissatisfaction expressed by the participants, we identified six challenges, detailed in the following paragraphs, that prevented participants from using \cbas to generate useful and satisfactory codebases.

\noindent\textbf{Missing and Blank Code.}
Participants were mostly dissatisfied by missing code---references to non-existent code, unimplemented code and files, or non-existent imports. These missing pieces ranged from missing variables and functions to missing entire files
We found that the \cba failed to generate the necessary files in 12 tasks. 
This included instances like a missing ``package.json" file required by Node.js (\P8) as well as instances where one file would try to import code from another one that did not exist (\P12).
We also observed some cases where \cba imported non-existent functions or referenced invalid attributes of existing classes.

In seven other tasks, the \cba left part of the code blank or generated completely blank code. As P11 experienced, ``it basically did not do anything. It just got us one read me file from GitHub, which tells us how to do the rest of the stuff. But it did not generate any code." Whereas, for P6, instead of the actual code implementation, the \cba left a comment such as ``Implement AJAX request to update the size." This frustrated participants, as \P9 said, ``If you're going to put in a function that's not implemented, then maybe you need to better specify that the function needs to be implemented!"

\noindent\textbf{Inadequate Communication.}
In 31/48 tasks, the participants failed to communicate in detail their needs and submitted partial prompts with unmentioned specifications. When we compared the requirements that participants listed verbally before writing down the prompt to the requirements expressed in the prompt, we found that participants frequently omitted requirements that they had previously articulated aloud.
In 20 out of 48 tasks, the written prompt failed to include at least one verbally stated requirement, with 13 of 16 participants making such omissions at least once. Participants identified these omissions when evaluating the generated codebases, e.g., P13 realized, ``maybe I should have specified `Oh! keep looping or keep checking for user input.'"
This gap suggests that prompt-writing imposes cognitive load, and that translating intentions into effective prompts needs further support. 

On the other hand, seven participants expressed that the generated code had bad variable names (P15), unnecessarily high cyclomatic complexity and often left out comments and explanations (P1, P6, P10, P12, P16), making it difficult to read (P10).
P6 was unable to even parse if the code generated was a database or function, stating, ``Is it some sort of database? I can't figure out! \ldots it would have helped if it did leave some comments." This demonstrates that \cbas need communicate their process and generated code better. 

\noindent\textbf{Ignored Context.}
Participants encountered \cbas ignoring existing code in 12 of the 24 edit tasks, making it the primary and sole contributor of ``poor maintainability" (P1,P2,P4,P8-P10,P12-P16).
In some cases, the \cbas even ignored the direction to use a specific piece of code.  P16 wrote ``Edit [the] calculator.py code to add [an] option to 'enter operation', called 'exponential'." 
Here, ``enter operation" referred to a string containing a list of operation options (``add", ``multiply", etc.), which the \cba did not edit or use. This suggests limited context parsing and selection in \cbas create incorrect code that participants found hard to maintain.

\noindent\textbf{Usage Instructions.}
In 10 instances, participants were confused about how to use the generated codebases. In some cases, the \cbas produced the correct code, but ``didn't tell [them] where to put the code" [P3]. In other cases, the \cbas produced jumbled, inconsistent parts of codebases that required reconstruction. P4 shared this experience, ``It seemed like it wasn't quite consistent\ldots\;It gave me like an app when I needed two different functions or something. It was confusing how to add it to the functions." In case of edit tasks, participants struggled to identify the differences, as P6 noted, ``they should also provide comments as to what changes have been made as compared to the previous code.''

\noindent\textbf{Partially correct or Unusable Code.}
In 17 cases, participants encountered  code that only partially works and meets their expectations or was completely incorrect or non-executable. This lead to logical issues with the generated codebase in 11 instances (P2,P5,P9-P11,13,P15) and syntax issues or runtime errors in six cases (P1,P5,P13,P16). P10 emphasized the challenges of partial or incorrect code stating, ``it [could] actually put someone in the wrong direction, right?" This misdirection frustrated participants, because ``[they] need to start over, need to do everything again, because [they] cannot use this code" (P12). Repairing these partially correct and unusable code often required additional effort, and calls for \cbas equipped with debugging and repairing assistance.

\noindent\textbf{Ignored Requirements}
In a few instances (3), we found that the \cba ignored requirements directly specified by participants. P1 described his experience stating, ``it ignored some of what I asked him to do" and P10 mentioned, ``it did not implement any of the semantics that I expressed very clearly." When \cbas ignored participants' specifications, it naturally resulted in missing functionality and incomplete code. This indicates that missing self-accountability checks in \cbas further add challenges for users.

We conducted a two-way ANOVA followed by post-hoc t-tests to explore how students mention code evaluation criteria differently. The ANOVA results showed that at least one of the factors differed significantly between students and developers ($F(1,11)=3.46, p=0.0004$). The post-hoc t-tests for each factor showed that “Correctness” was mentioned significantly more by developers ($p = 0.034$), while other differences were not statistically significant.

\keyfinding{
Six root causes explain dissatisfaction: incomplete code, incorrect code, unmentioned or ignored requirements, ignored context, and missing guidance—spanning both technical and interaction failures.
}

\section{\textbf{RQ4.} What are the barriers to integrating \cbas into developer workflows?}
Building upon the challenges of using \cbas efficiently, our interviews revealed underlying reasons why participants may be reluctant to incorporate these tools into their workflows.
Participants identified significant barriers to adopting \cbas, citing that they have limited capabilities, requiring effortful and uncontrollable interactions, which isn't necessarily faster than coding alone, and can bear legal or privacy challenges. 

\subsubsection{Limited Capability}
Many participants questioned the functional adequacy of \cbas, especially for tasks requiring complex reasoning, contextual understanding, or integration with existing codebases. P8 questioned their competence by saying, ``It has very, very simple answers," adding that they're ``not ... as smart as a human." 
Concerns also extended to the quality of the code.
P13 emphasized, ``It's pretty important to have code that runs... It shouldn't create... code that doesn't run in the first place, and P16 criticized it for producing ``non-compiling" code.  
Architectural shortcomings were pointed out by participants, as P1 explained,  ``[the codebase] doesn't really work well enough unless I know what the actual architecture should be." 
These issues eroded participants' trust in the tool’s ability to handle nuanced or non-trivial tasks in the long run.

\subsubsection{Effort Overhead}  
Five participants (P1, P5, P6, P8, P9, P16) expressed concerns that using \cbas often imposed additional effort instead of reducing it. As P1 exclaimed when evaluating the codebase, ``we will have to make a lot of changes!"
P9 noted, ``it's [as] difficult [as] debugging somebody else's code." This was partially due to the lack of structure or explanation in the codebases, as P6 elaborated, ``[they are] not giving me any documentation or comments." 

Apart from comprehending code, participants also agreed on the effort overload from elaborate prompting. P16 shared that specifying file context ``at all points was labor-some." Manual file selection (P5) and lack of support or documentation (P6, P8) further added to the friction.
For some, this overhead canceled out reduction in cognitive effort that \cbas provided.

\subsubsection{Lack of Control}  
A recurring issue was the unpredictability of the AI agents' behavior.  
Participants (P2, P13, P14) found it unclear what \cbas could do or how specific their prompts needed to be.  
P14 pointed out the challenges with writing prompts since, ``it's not very clear how the prompt should be written." P13 emphasized, ``I think it's tricky to know what it can and can't do", underlining how the uncertainty about \cbas' capabilities further reduced the feeling of control.  
This ambiguity made it hard for participants to feel in control of the tool, hindering its usability for tasks requiring precision and reliability.

\subsubsection{No Time Gains}  
Despite anecdotes and advertisements about increased productivity that comes with \cbas, several participants (P2, P8, P14, P16) observed that using \cbas did not significantly speed up their workflows.  
As P2 put it, ``if I have to intervene anyway \ldots might as well just do it myself."  
Similarly, P14 explained, ``if you're implementing an algorithm, you might have to like, sit and explain the algorithm, whereas it might just be easier to type it out yourself."  
This suggests that participants felt the time required to prompt, clarify, and revise \cba output often offset any time saved for some tasks. 

\subsubsection{Legal and Privacy Concerns}
Finally, a few participants raised concerns around code licensing and data confidentiality issues of \cba generated code. Such concerns are commonly reported in any AI-generated code~\cite{akbar2023ethical,stalnaker2024developer}. P14 pointed out ``privacy'' as their main concern, and P9 added that they experienced first-hand ``concerns of shipping code using LLMs" due to legal ambiguities in how training data might affect code ownership.  
Such concerns highlight the need for organizational policies and clearer legal frameworks before \cbas can be widely adopted in professional contexts.

\keyfinding{
Barriers to adopting \cbas in development workflows range from their limited capability to handle complex scenarios (50\%) and the extensive effort (31\%) and time (25\%) needed to generate satisfactory code from \cbas. Participants also identified having a lack of control over \cba output and legal repercussions of using \cba generated code as barriers to adopting the tool.
}

\begin{table}[h]
\centering
\vspace{-1em}
\caption{Selected repository-level code-generation tools}
\vspace{-0.5em}
\begin{tabularx}{\columnwidth}{p{0.2cm} X}
\toprule
\textbf{Type} & \textbf{Tools} \\
\midrule
\IDEicon &
    Copilot~\cite{ghcopilot_doc}, Cursor~\cite{cursor_agent_docs}, Cline~\cite{cline_doc}, 
    Windsurf~\cite{windsurf_doc}, Softgen~\cite{softgen_doc}, Pear~\cite{pear_doc} \\
\CLIicon &
    GPT-Engineer~\cite{lovable_doc}, Aider~\cite{aider_doc}, Codebuff~\cite{codebuff_doc} \\
\SDEicon &
    GitHub Workspace~\cite{ghworkspace_doc}, Devin~\cite{devin} \\
\WEBicon &
    Databutton~\cite{databutton_doc}, Replit~\cite{replit_doc}, Base44~\cite{base44_doc},
    Qodo~\cite{qodo_doc}, Srcbook~\cite{srcbook_doc}, Pythogara~\cite{pythagora_doc},
    Bolt~\cite{bolt_doc}, V0~\cite{v0_doc}, Webdraw~\cite{webdraw_doc}, Tempo~\cite{tempo_doc} \\
\bottomrule
\end{tabularx}
\vspace{-0.5em}

\label{tab:rlcg-toollist}
\end{table}
\section{Discussion}
Following our findings in the earlier sections, we look into a set of existing \cbas and compare their capabilities.
This way we could find out how \cba capabilities are addressing software developers' needs and what is left to be addressed.
We conclude by offering recommendations to improve \cbas, aiming to address their interaction challenges with developers and facilitate integration into programming workflows.

\subsection{Current \cba Capabilities and Design Opportunities}
\label{sec:background-rlcg}

In the past year, numerous new \cbas have been launched with diverse functionalities, interaction models, verification strategies, and integration methods. Which of the \cbas are effective and can generate useful codebases? Do their features mitigate the \textit{challenges} and \textit{barriers} identified in our results? 

To compare user challenges and barriers with current \cba capabilities, we analyzed a representative set of 21 widely used \cba tools designed for different users and interaction modes (Table \ref{tab:rlcg-toollist}).
Based on interaction modes, these \cba fall into four categories:
\IDEicon \emph{IDE Agents} are tightly integrated into developers’ Integrated Development Environments (IDEs), like Cursor and Copilot. 
\CLIicon \emph{CLI Agents} operate primarily through the Command Line Interface (CLI), aimed at helping build prototypes or scaffolding, primarily by technical users. E.g., GPT-Engineer.
\SDEicon \emph{SDE Agents}, such as Devin and GitHub Workspace, which are designed for organizational or team-level workflows to function as collaborative agents.
\WEBicon \emph{Web agents} require no programming knowledge and allow users to interact with natural language, targeting low-code or no-code access to design and automation. A popular example is Webdraw, which lets users convert their sketches into websites.

Two researchers began with an exhaustive inventory of individual features advertised by these tools and conducted an open‑coded thematic analysis, iteratively grouping similar features into broader capability dimensions. 

\vspace{0.1cm}
\textbf{R‑CTX}\textit{: How the system retrieves context.}
\\
Effective code generation in real-world codebases requires rich contextual information to account for dependencies between files, functions, and modules~\cite{hai2024impacts}.
However, despite this need for context, LLMs often struggle to retrieve and utilize relevant code context effectively~\cite{jimenez2023swe,li2023large,an2024make}.
\WEBicon{} and \CLIicon{} agents, which focus on building codebases from scratch, either \textit{lack this capability} entirely (e.g., Base44, Webdraw) or append entire files without discrimination (e.g., GPT-Engineer). In contrast, \IDEicon{} and \SDEicon{} agents \textit{can retrieve relevant context automatically}, using symbol indexing or embedding repository content, which aligns with retrieval-augmented generation (RAG)~\cite{gao2023retrieval}. These agents also allow users to tag important files (e.g., Devin, Pear, Cline) or highlight them (e.g., Copilot) to prioritize their inclusion as contextual input.

\underline{Opportunities:} We found that participants faced challenges when \cbas ignored the code context (RQ3 challenge: ignored context), which required them to spend more time and effort on achieving a satisfactory codebase (RQ4 barrier: limited capability, effort overhead, no time gains). \cba developers should focus not only on retrieving context, but also on identifying the optimal context for a prompt and using the context dynamically to generate satisfactory codebases.

\vspace{0.1cm}
\textbf{PL-SHARE}\textit{: How the system shares its plan with the user.}\\
The ability to plan is a critical capability for codebase agents, as it defines the scope and sequencing of the multiple modifications required to complete a task~\cite{jiang2024self}. As a result, self-planning has become a central feature across many codebase agents. Some agents, such as GPT-Engineer and GitHub Workspace, explicitly share their plans with users prior to code generation.
This supports human-in-the-loop interaction~\cite{huff1988plan,pollock1999planning}.
In contrast, other agents either do not expose their planning process at all (e.g., Webdraw and V0), or do not solicit user feedback on the plan (e.g., Aider).
The latter approach is more common in systems like \WEBicon, where the target user population may lack the expertise to effectively evaluate or influence the agent’s plan.

\underline{Opportunities:} Our participants struggled when \cbas were not transparent in communication about what they planned to implement (RQ3 challenge: inadequate communication), or when the \cbas didn't provide a detailed plan on how to implement the generated codebase (RQ3 challenge: usage instructions). \cbas must explain their implementation plan to the user and ask for clarification/modification where feasible. Agents already doing so should strive for collaborative planning with users to allow them more control of the generated codebase (RQ4 barrier: lack of control). \cbas can also aim to build code hierarchically, starting with proposing the scaffold of a codebase, then filling in more granular parts down to the function level. Similar hierarchical approaches has been taken by Yao et. al. to enhance complex capabilities of LLMs such as problem solving and reasoning~\cite{yao2023tree}.

\vspace{0.1cm}
\textbf{S-VER}\textit{: How the system verifies its outputs.}\\
LLM-generated code is frequently found to be incorrect or error-prone~\cite{liu2023your,vasconcelos2022generation}.
To address these challenges, prior work has explored how traditional software verification techniques can be integrated with LLM-based systems~\cite{tihanyi2023formai,mitchell2025can}.
Some other \cbas have adopted self-reflection as a means of verification~\cite{robeyns2025self}. Tools like Cursor and Devin implement iterative self-verification cycles in which the agent examines its own output, identifies errors, and revises the code accordingly~\cite{devin,cursor_agent_docs}.
Others, like Pythagora, employ multi-agent frameworks where specialized agents are responsible for detecting potential faults and relaying feedback to the primary generation model.
These self-refinement processes allow for autonomous correction without requiring human intervention~\cite{jiang2024self,sadra-trust}.

\underline{Opportunities:} Participants struggled with \cbas producing partially correct code (RQ3 challenge: Partially correct or Unusable Code) and or when code behavior deviated from specifications (RQ3 challenge: Ignored requirements). They report difficulties in debugging and correcting bad code, expressing the need for assistance. \cbas should aim to not only self-correct, but also provide affordances and instructions for users (especially non-technical users) to verify the outputs through verification modules and explanations.

\vspace{0.1cm}
\textbf{PROACT}\textit{: How the system proactively adds features.}\\
Several code generation tools—particularly those in \WEBicon and \SDEicon contexts---proactively suggest or apply code changes even when users have not explicitly requested them. This behavior may stem from the sycophantic tendencies of LLMs, which are trained to optimize for user satisfaction by anticipating and aligning with perceived user intent~\cite{sharma2023towards}.
While such an initiative is often valued by product stakeholders who prioritize rapid prototyping and code visibility~\cite{ademvibe}, it is perceived as invasive by users as it further reduces their control over the output produced by \cbas (RQ4 barrier: lack of control). Prior research has highlighted the potential risks associated with unsolicited code modifications, including challenges in maintainability~\cite{maes2025gotchas}, reduced user control~\cite{maes2025ensuring}, and the possibility of introducing unsafe or unintended behavior~\cite{sadra-trust}.

\underline{Opportunities:} Proactive code generation can function as a Trojan horse--offering convenience and apparent progress, while obscuring long-term implications on quality and developer agency.
Future work could focus on reducing LLM's sycophantic tendencies \cite{sharma2023towards} that produce unsolicited changes. Promising approaches include multi-agent LLM frameworks \cite{ishibashi2024self} that audit suggestions and ensure explicit developer approval. Additionally, granular interaction loops structuring prompt-code cycles to incrementally validate features can ensure generated outputs align strictly with user requests \cite{sadra-trust}.

\vspace{0.1cm}
\textbf{PR}\textit{: How the system proposes the incoming change.}\\
\cba tools vary in how they present proposed changes to users.
Agents in \WEBicon and \CLIicon settings tend to overwrite the entire codebase (e.g., GPT-Engineer), mirroring how less-experienced developers frequently interact with AI-generated suggestions by copy-pasting whole codebases~\cite{hamer2024just}. In contrast, agents embedded in \IDEicon and \SDEicon workflows (like Devin) more commonly adopt pull request mechanisms, reflecting norms in open-source~\cite{rahman2014insight} and industrial development~\cite{lenarduzzi2021does}. This allows developers a sense of control, reduce the barrier of loss of agency (RQ4 barrier: lack of control).

Some other \cbas (e.g., Cursor, Cline) present their proposed changes as ``diff'' files, highlighting modifications relative to the current codebase~\cite{mackenzie2002comparing}.
These practices position the agent as a collaborative participant in the development process—akin to a human team member~\cite{defranco2017review} or a partner in a co-programming setting~\cite{hannay2009effectiveness}.

\underline{Opportunities:} \cbas must adequately communicate any proposed changes and provide guidance and detailed instructions on how to implement the changes in the codebase. While ``diff" or ``pull requests" can help technical users, participants expressed the need for stepwise explanation of the proposals, using template codes, and rationale behind the proposals. 

\vspace{0.1cm}
\textbf{TEACH}\textit{: Whether the system tries to teach or onboard users with the generated code.}\\
Current tools vary in their support for educational value.
\WEBicon agents typically present only final outputs, offering minimal opportunity for user comprehension.
Some tools such as Devin and GPT-Engineer, provide high-level summaries of code changes.
Even among \IDEicon tools--where the interaction is more direct--explanations are often only surfaced upon request, such as triggers like \verb|\explain| in Copilot.

\underline{Opportunities:} We observed users facing challenges due to the absence of guidance (RQ3 challenge: usage instructions). Prior work has also shown that developers are more likely to trust and adopt AI-generated code when they understand the underlying changes~\cite{sadra-trust,johnson2023make} (RQ3 challenge: inadequate communication).
Features like interactive walkthroughs and explanations can play a dual role--facilitating learning while also improving maintainability~\cite{sadra-trust}.

\vspace{0.1cm}
\textbf{ASK}\textit{: Whether the system clarifies ambiguities.}\\
Some tools, such as GPT-Engineer, attempt to ask clarification questions when prompted requirements are ambiguous.
However, many agents avoid querying users, reflecting on the priority of smooth interaction over precision.
To address this, Pythagora introduces a dedicated agent, \emph{SpecWriter}, which helps clarify the scope of a task during prompt formulation.
While these mechanisms reduce the burden on users, they can also reduce control and lead to unintended outcomes.

\underline{Opportunities:} Our study showed that developers often missed requirements while writing the prompt. Even when specifying requirements, \cbas sometimes ignores them (RQ3 challenge: ignored requirement).
\cbas asking questions that can lead developers to reflect on their requirements could help retrieve those forgotten items from the developers' memory.
Clarification during interaction with AI is known to be impactful on the quality of the output~\cite{mu2023clarifygpt}.
Sabouri et. al. introduced the notion of double-sided clarification for improving alignment, where both the developer and the AI system actively seek to refine the problem specification~\cite{sadra-trust}.
\vspace{-0.1cm}
\subsection{The Art of Prompt Crafting.}
Due to the inherent uncertainty regarding details when writing code, many participants in our study struggled with crafting ``good'' prompts that result in their desired code. This process becomes more challenging with \cbas with a larger generation scope. As each prompt results in more lines of code, it makes it difficult to iteratively adjust the prompts afterwards. 

Our findings suggest that a more guided prompting process may support users in expressing their intent adequately.
Rather than requiring a well-formed prompt upfront, \cbas could scaffold prompt development by asking for a high-level description and follow up with clarifying questions and suggesting refinements.
Such a system might include a conversational pre-prompting stage that elicits goals, constraints, and references before translating them into a structured prompt.
By offloading the burden of prompt construction, guided prompting could reduce cognitive load and frustration and improve alignment between user expectations and code generation.
\vspace{0.05cm}
\section{Threats to Validity}
\emph{Internal Validity.} Like any other user study, our participants may have been influenced due to observer bias (Hawthorne effect~\cite{sedgwick2015understanding}).
We mitigated this by creating a study with minimal intervention, and instead asked them to verbalize their thought process (think-aloud protocol~\cite{jaaskelainen2012think} is a common tradeoff in HCI research that enables insight into participants’ reasoning with minimal disruption). To reduce learning effects in participants completing three tasks in a single session (48–85 minutes), task order was randomized and counterbalanced.

\emph{External Validity.} Our participants were US-based, which may limit cultural or geographical generalizability. Nonetheless, the tools and workflows studied are internationally standard. The small sample size of our participants restricts broad generalization of our findings, but offers deep and rich qualitative insights and is consistent with prior work in the venue~\cite{infoSeek}.
While we counterbalanced four other factors, we were unable to control for participants’ prior frequency of AI tool usage, since recruitment and experimentation were conducted in parallel and the distribution was not known a priori. 
This is unlikely to significantly impact our results, as we do not directly compare interactions across \cbas. 
However, we acknowledge this as a potential limitation and recommend future work more systematically account for prior exposure to AI tools.
While we were limited by time and feasibility to refrain from using complex and collaborative tasks, our designed tasks reflected realistic create and edit cases.
Additionally, we used controlled tasks with artificial codebases to standardize task complexity and tooling, facilitating systematic comparison of prompting behavior and \cba effectiveness.
While aligned with prior research~\cite{VLHCC2020,CHI2014}, this approach does not fully capture developers’ naturalistic interactions with familiar projects.
Future field studies are needed to explore this aspect.
Finally, we focused on participants’ first prompt to isolate initial \cba alignment challenges, keep sessions consistent, and limit fatigue. 
We recognize this reflects only part of real-world iterative prompting and recommend future studies on multi-round prompt refinement.

\emph{Construct Validity.} To reduce bias in qualitative coding, two independent coders validated the prompt and factor categorizations. To ensure objectivity and confidence in interpreting user complaints (e.g., “incomplete code”) and attribution of actions (e.g., omitted requirements), we cross-referenced prompts with generated code and explicit participant elucidations and used inductive coding with high inter-rater agreement ($\kappa = 0.87$)

\emph{Conclusion Validity.} Our reported correlations (e.g., between prompt style and satisfaction) have a small size, are based on a small sample, and are meant to be exploratory. Our study captured only short-term usage in a lab setting and did not assess longer-term adoption or evolving practices.
Given this scope, our goal was to identify initial usability challenges; future longitudinal work is needed to understand sustained use.
  
\section{Conclusion}
We conducted a study with 16 developers and student programmers to explore the challenges and opportunities of codebase-level assistants. Our analysis revealed diverse prompting styles, frequent omission of verbally stated requirements, and moderate satisfaction with \cba outputs, driven by functional correctness, executability, and clarity. Participants raised concerns about model limitations, loss of control, added effort, and privacy risks, which hinder adoption. To address these, we recommend richer guided prompting, transparent planning, iterative generation with validation, and clearer presentation of changes. These steps can better align AI codebase assistants with professional workflows, making them more reliable, efficient, and user-centered.

\bibliographystyle{IEEEtran}
\bibliography{references}

\end{document}